\documentclass{article}

\usepackage[english]{babel}

\usepackage[letterpaper,top=2cm,bottom=2cm,left=3cm,right=3cm,marginparwidth=1.75cm]{geometry}

\usepackage{amsmath}
\usepackage{graphicx}
\usepackage[colorlinks=true, allcolors=blue]{hyperref}
\usepackage{tabularx}
\usepackage{booktabs}

\usepackage{lineno}

\title{The Computational Foundations of Collective Intelligence}

\author{
  Charlie Pilgrim$^*$\\
  University of Leeds, UK
  \and
  Joe Morford\\
  University of Rochester, USA
  \and
  Elizabeth Warren \\
  Johns Hopkins University, USA
  \and 
    Mélisande Aellen \\
  University of Rochester, USA
  \and
  Christopher Krupenye \\
  Johns Hopkins University, USA
  \and 
  Richard P Mann \\
  University of Leeds, UK
  \and
  Dora Biro \\
  University of Rochester, USA
  \\\\
  $^*$Email: \texttt{c.p.pilgrim@leeds.ac.uk}
}

\begin{document}
\maketitle

\begin{abstract}
    
    Why do collectives outperform individuals when solving some problems? Fundamentally, collectives have greater computational resources with more sensory information, more memory, more processing capacity, and more ways to act. While greater resources present opportunities, there are also challenges in coordination and cooperation inherent in collectives with distributed, modular structures. Despite these challenges, we show how collective resource advantages lead directly to well-known forms of collective intelligence including the wisdom of the crowd, collective sensing, division of labour, and cultural learning. Our framework also generates testable predictions about collective capabilities in distributed reasoning and context-dependent behavioural switching. Through case studies of animal navigation and decision-making, we demonstrate how collectives leverage their computational resources to solve problems not only more effectively than individuals, but by using qualitatively different problem-solving strategies.  
\end{abstract}

\section*{Significance Statement}

There are many ways in which groups can outperform individuals, from the wisdom of crowds to cultural learning and specialisation. However, explanatory approaches remain fragmented and siloed, each focused on a subset of behaviours. We present a framework that unifies these approaches by analysing how collectives differ from individuals in their computational resources and constraints. Our approach not only provides a shared conceptual language for interdisciplinary work, but also points to novel predictions in emergent collective reasoning and collective adaptation. While our focus is on animal groups, the principles described here can be applied to human societies, neural circuits, and other biological systems. 

\section{Introduction}

Collectives outperform individuals when solving problems in a diverse range of contexts: ant colonies choosing a nest site \cite{franks2003strategies, franks2002information, pratt2002quorum}, fish navigating their environment \cite{berdahl2013emergent, grunbaum1998schooling, couzin2018collective}, meerkats keeping a lookout for predators \cite{clutton1999selfish}, and humans accumulating knowledge and skills \cite{tennie2009ratcheting, tomasello1993cultural, boyd1996culture}. What connects these forms of collective intelligence? 

Existing integrative efforts, taken alone, are compelling and point to important general principles. However, the diversity of explanatory approaches reveals that none are truly comprehensive, and research remains siloed across disciplines. For example, early insights in Condorcet's jury theorem \cite{de1785essai} and the wisdom of the crowd \cite{francis1907vox} highlight the statistical benefits of aggregating opinions to improve judgement accuracy \cite{bernoulli1713ars, de1738doctrine, de1810memoire}. Relatedly, collective sensing relies on a different form of information aggregation, with individuals sharing distributed sensory information to more effectively respond to the environment  \cite{berdahl2013emergent, couzin2007collective, pulliam1973advantages}. In contrast, cultural learning relies on the transmission of knowledge across generations \cite{tomasello1993cultural}, generating long-term persistence of memory \cite{sasaki2017cumulative} and the potential ratcheting up of collective capacities \cite{boyd1988culture, tennie2009ratcheting, tomasello1993cultural, boyd1996culture}. Division of labour requires yet another distinct explanatory approach, with improved group performance through complementary actions and role specialisation \cite{smith1776inquiry, becker1992division, anderson2001teams, west2015major}. From a complexity science perspective, Sumpter \cite{sumpter2006principles} calls for a focus on how collective behaviour emerges from interaction rules between individuals, while Galesic et al. \cite{galesic2023beyond} go further by asking how these interaction rules adapt to changing problem contexts. Underpinning much of our modern understanding of collective intelligence is Couzin's approach to bridging empirical behaviour with theoretical collective intelligence \cite{katz2011inferring, couzin2003self, couzin2002collective, couzin2005effective}, including drawing parallels between animal groups and neuronal processes \cite{couzin2009collective}. Taken together, these varied explanatory approaches call for a synthesising framework. 

\begin{figure}[ht]
\centering
\includegraphics[width=.9\linewidth]{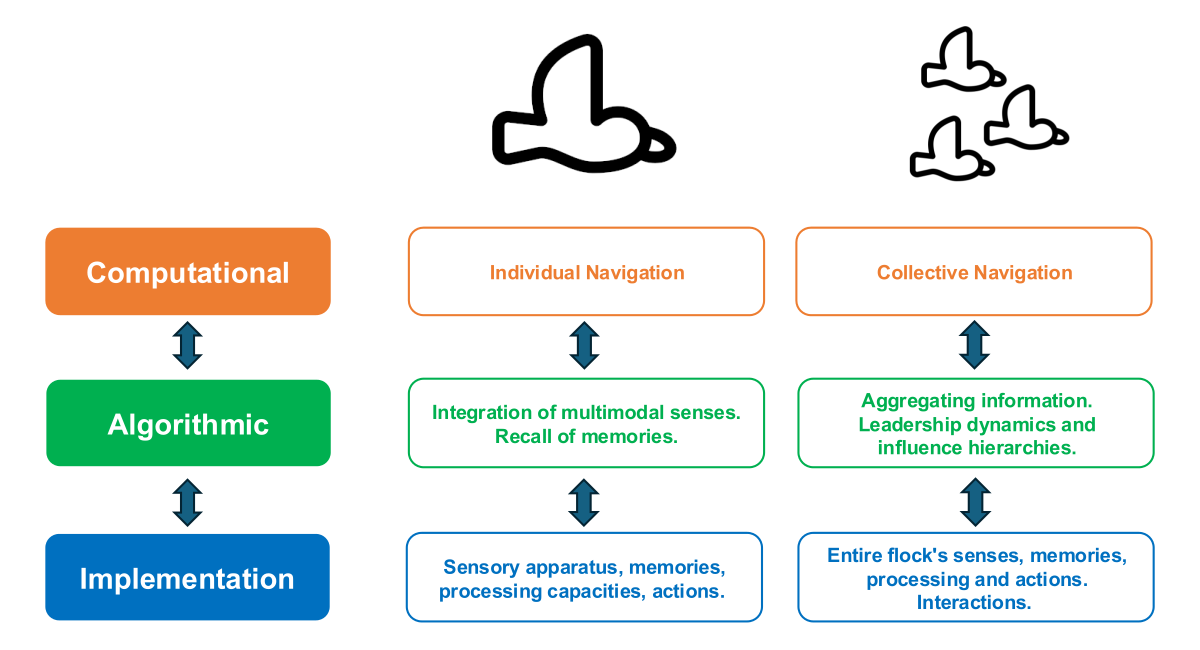}
\caption{Analysis of problem solving across Marr's levels of analysis (left column) for individual homing pigeons (central column) and flocks (right column). Collectives have greater resources and different structural constraints, enabling different algorithms, and ultimately solving the problem of navigating home more efficiently.}
\label{fig:pigeon}
\end{figure}

Our contribution is to show that these disparate accounts are aspects of a single unifying principle: collectives differ from individuals in their computational resources and constraints. For example, a solitary homing pigeon processes multimodal sensory information \cite{wallraff2004avian, schmidt1990sun} along with memories of landmarks \cite{biro2007pigeons} in order to choose which direction to fly (Figure \ref{fig:pigeon}). In contrast, a flock of pigeons has more resources at their disposal: the entire flock's senses and memories, the internal processing within each pigeon as well as interactions between birds \cite{sasaki2018personality, nagy2010hierarchical}, and options to fly together or apart \cite{biro2006compromise}. Fundamentally, it is through leveraging these enhanced collective resources that a flock of pigeons is able to solve the problem of navigating home more efficiently than an individual \cite{sasaki2022empirical}.

\begin{figure}[ht]
\centering
\includegraphics[width=.9\linewidth]{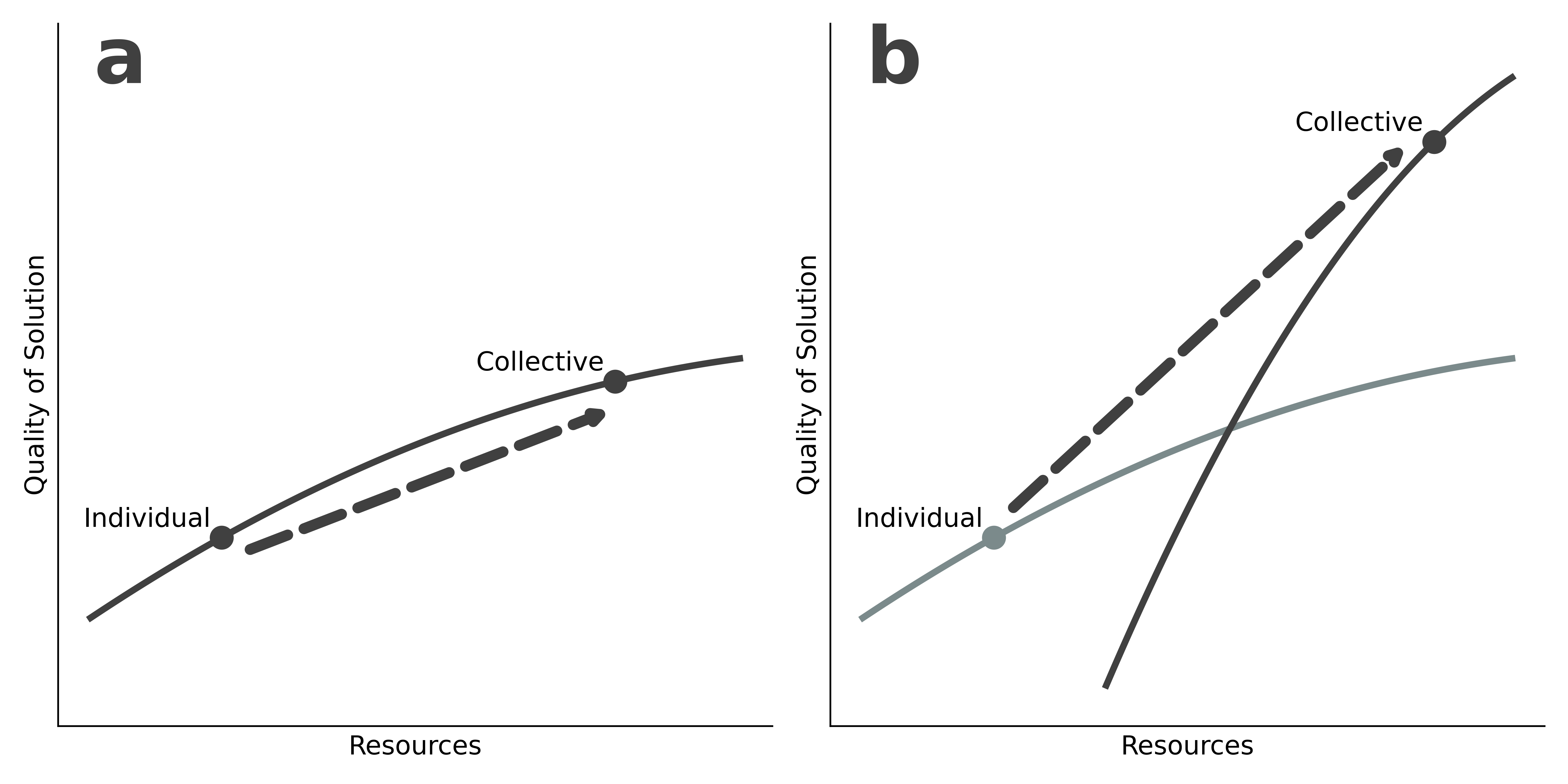}
\caption{Improvements in the quality of solutions available to collectives over individuals. a) Quantitative improvements involve a collective solving a problem in similar ways to individuals but with greater resources, moving along the same resource-quality curve. b) Qualitative improvements involve new forms of problem-solving at the collective level that implicate a different resource-quality curve.}
\label{fig:resource_quality_curves}
\end{figure}

This resource-focussed view aligns with the idea of resource rationality: behaviour can be analysed as a constrained optimisation that takes into account not only the quality of problem-solving but also resource costs \cite{griffiths2015rational} (see also bounded \cite{simon1955behavioral}/ecological\cite{todd2012ecological, gigerenzer2004fast}/adaptive \cite{haselton2009adaptive} rationality). On the one hand, collectives have greater resources than individuals -- more sensory information, greater capacities in both memory and processing, and more ways to act. On the other hand, collectives face additional constraints because those resources are distributed across autonomous individuals \cite{velez2023humans}, raising challenges in coordination \cite{steiner1972group, brooks1995mythical, malone1994interdisciplinary, krafft2018levels} and cooperation \cite{steiner1972group, krafft2018levels}. Overall we can say that collectives face different resource-quality tradeoffs than individuals. While in some contexts these tradeoffs cause collective failures (e.g. groupthink \cite{janis1973groupthink}, mutual defection \cite{olson1965logic}, spiralling costs \cite{brooks1995mythical}), in other contexts collective resource advantages lead to quantitative improvements in problem-solving (e.g. accuracy, speed; Figure \ref{fig:resource_quality_curves} \textbf{a}), as well as qualitative improvements whereby the collective solves problems in ways that individuals cannot (Figure \ref{fig:resource_quality_curves} \textbf{b}). 

In order to connect physical resources to problem-solving, we are inspired by Marr's levels of analysis. Specifically, we take from Marr the idea that information processing systems can be analysed at three separate but connected levels \cite{marr2010vision}: i) Computational, the real-world problem being solved; ii) Algorithmic, how information is represented and processed to solve the problem; iii) Implementation, the physical reality of implementing those algorithms (e.g., in neural substrate or physical hardware). While Marr emphasised a top-down analysis starting with the real-world problem being solved, we invert this approach in order to ask how collective resources (Implementation) lead to information processing capacities (Algorithmic) and ultimately enhanced problem solving (Computational). 

In what follows, we begin by quantifying collective and individual resources, and accounting for coordination and cooperation challenges. We then describe how collective resources can translate to algorithmic capacities, explaining traditional forms of collective intelligence in the wisdom of the crowd \cite{surowiecki2005wisdom, francis1907vox, simons2004many, page2008difference}, collective sensing \cite{couzin2009collective, lima1995back, hein2015evolution}, division of labour \cite{robinson2009division, becker1992division, szathmary1995major}, cultural learning \cite{tennie2009ratcheting, tomasello1993cultural, boyd1996culture} --- as well as supporting emerging research in collective adaptation \cite{galesic2023beyond}, collective learning \cite{biro2016bringing, kao2014collective}, inference \cite{arganda2012common, marshall2009optimal} and switching between deliberation strategies \cite{seeley2012stop, strandburg2015shared}. To demonstrate the utility of this framework to understand empirical behaviour, we then present a series of real-world case studies of problem-solving in animal groups.  Finally, we explore how our framework opens up new questions, from collective reasoning and rationality to the applicability of these principles to neural processes. 

\section{Collective Resources (Implementation Level)}

In order to quantify resources, we must first define the dimensions along which resources can vary. We are influenced by the wide literature on formal models of collective intelligence \cite{sumpter2006principles, francis1907vox, berdahl2013emergent, pulliam1973advantages, couzin2007collective, tomasello1993cultural, tennie2009ratcheting, anderson2001teams, galesic2023beyond, couzin2003self, couzin2002collective, couzin2009collective}. We also draw on models of cognition in the ACT-R \cite{anderson2014atomic, anderson2013architecture} and SOAR \cite{newell1994unified} cognitive architectures, connectionist paradigms \cite{mcclelland1995there, rumelhart1986general} and frameworks for extended \cite{clark1998extended} and embodied \cite{varela1991embodied} cognition; as well as fundamental models of computation in Turing machines \cite{turing1936computable} and the Von Neumann architecture \cite{von1993first}. We identify four core resource dimensions that capture the computational process from perception to action and apply to both individuals and collectives (Table \ref{tbl:hardware}): 

\begin{enumerate}
    \item Sensory Information: Inputs received from the environment.
    \item States: The system configuration, including memory and location. 
    \item Processes: The functions available to process information within the system. 
    \item Actions: Outputs and final states. 
\end{enumerate}

\begin{table}[ht]
\centering
\small
\renewcommand{\arraystretch}{1.2}
\begin{tabularx}{\textwidth}{l >{\raggedright\arraybackslash}X X X}
\toprule
\textbf{Dimension} & \textbf{Individual} & \textbf{Collective} \\
\midrule
Sensory Information & $I_i$ & $\sum I_i$ \\
States (Memory and Location)     & $S_i$ & $S_c \times \prod S_i$ \\
Processes      & $F_i$ & $F_c + \sum F_i$ \\
Actions     & $A_i$ & $\prod A_i$\\
\bottomrule
\end{tabularx}

\caption{Resource capacities of individuals and collectives. The total sensory information received by the collective is the combined individual sensory information, $\sum I_i$. The collective state space is made up of collective level state (e.g. group structure) as well as the cross product of individual states, $\prod S_i$. Collectives can process information through inter-individual processes, $F_c$, as well as individual processing capacities, $\sum F_i$. The collective action space is the cross product of individual action options, $\prod A_i$. Note that the symbols $+$ and $\sum$ denote the total resources across the collective (not simple addition), $\times$ and $\prod$ indicate the Cartesian product.}
\label{tbl:hardware}
\end{table}

\begin{figure}[ht]
\centering
\includegraphics[width=.9\linewidth]{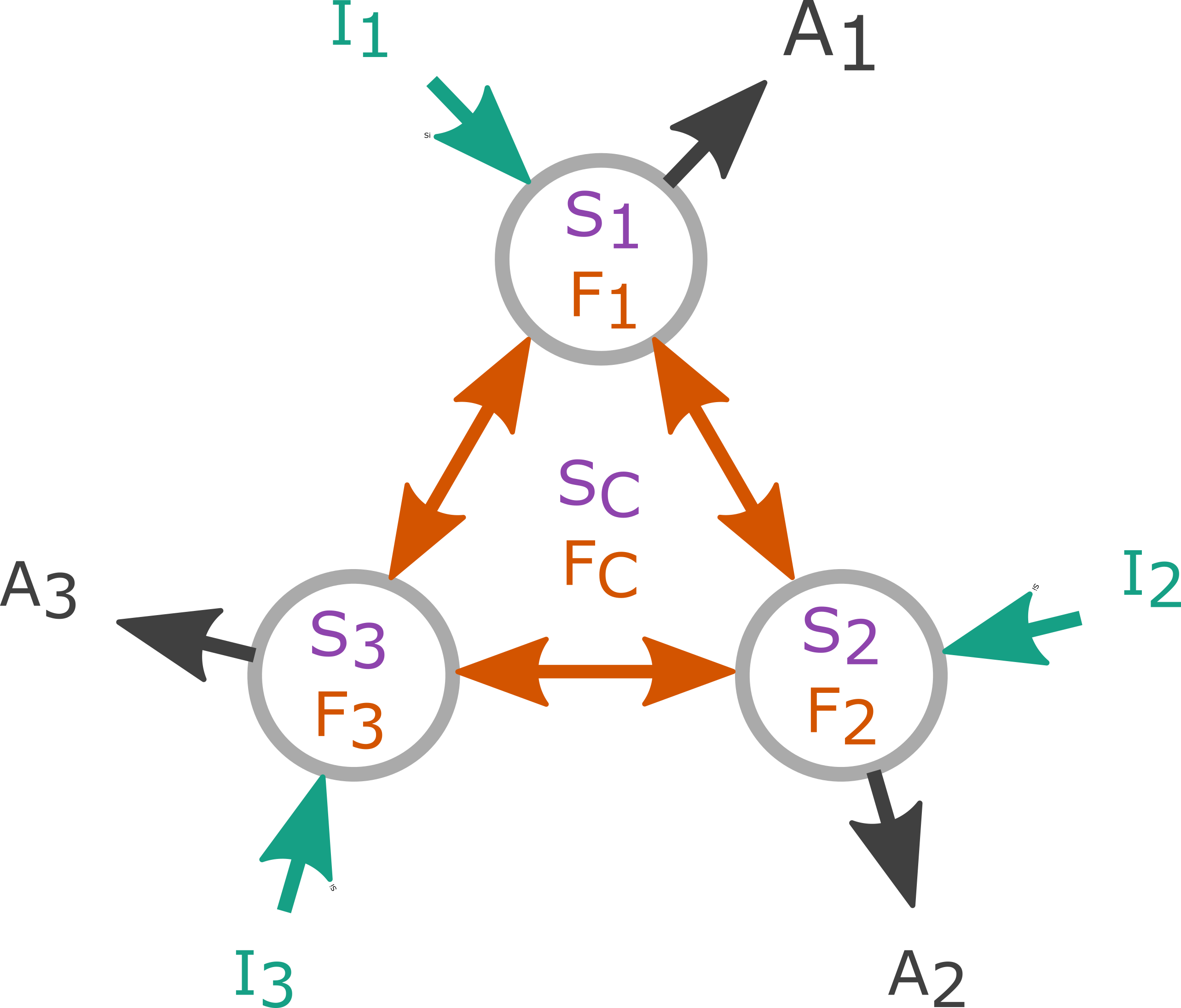}
\caption{Information processing schematic showing the structure of sensory information (green), states (purple), processes (orange) and actions (black) of a collective of three individuals. Each individual, $i$, receives sensory information, $I_i$, has a set of possible states $S_i$, a set of processes, $F_i$, and a set of output actions, $A_i$. The collective overall receives the sensory information of all individuals, $I_1 + I_2 + I_3$. The collective has access to processes arising from inter-individual interactions as well as the internal processes of each individual, $F_C + F_1 + F_2 + F_3$. The collective state space comprises all combinations of how individuals relate to each other and their internal states, $S_c \times S_1 \times S_2 \times S_3$. The collective action space is given by all possible individual actions, $A_1 \times A_2 \times A_3$. }
\label{fig:schematic}
\end{figure}

We quantify the hardware resources of individuals and collectives in Table \ref{tbl:hardware}. For clarity of comparison we focus only on computational resources embodied in agents and their interactions (Figure \ref{fig:schematic}). Both individuals and collectives also have access to environmental computational resources, such as external memory in pheromone trails \cite{grasse1959reconstruction, deneubourg1990self}. This form of extended or distributed cognition \cite{clark1998extended, hutchins1995cognition} enlarges the resources available and can play an important role in both individual and collective problem solving. We will return to this when discussing algorithms.

Note that the collective resources are not necessarily fully available to solve problems as a consequence of constraints and challenges in coordination and cooperation \cite{steiner1972group}, which we discuss in the next section. 

\subsection{Sensory Information}

For a collective, the sensory information received is the total sensory information received by individuals \cite{davidson2021collective, pulliam1973advantages}, $\sum I_i$. There can be redundancy in sensory information, with varying levels of correlations between the things that are being sensed \cite{kao2019modular} and the errors in those perceptions \cite{simons2004many, list2009independence}. 

\subsection{States (Memory and Location)}

The state of a system captures its complete configuration at a moment in time. State encompasses all forms of memory, physiological state, and location in physical or other spaces. The collective state space includes the state space of each individual, $\prod S_i$, as well as states that can be encoded in group structure, $S_c$, such as spatial configuration \cite{couzin2002collective} or interaction network topology \cite{biro2016bringing}. 

The state space of a collective, $S_c \times \prod S_i$, is exponentially larger than the state space of an individual (Table \ref{tbl:hardware}). For example, a single pigeon might be described by its location, stored information in memory, and biochemical status. To describe a flock of pigeons we would need the combined states of all pigeons and their relationships to each other.

\subsection{Processes}

Processes are the mechanisms by which information can be transformed. Information can be processed internally by each of the individuals within a collective, $\sum F_i$, as well as through inter-individual actions, $F_c$ (Table \ref{tbl:hardware}). These processes represent the primitive computational operations available to the system (which can be defined in different ways depending on the granularity of analysis). In the Collective Representations and Algorithms section we will discuss collective-level information processing, which emerges through the organisation and combination of the lower level processes defined here with other resources.  

\subsection{Actions}

The collective action space comprises all possible configurations of individual actions, $\prod A_i$. This action space scales exponentially with the number of individuals, with some constraints that limit the ability to take individual actions contemporaneously, e.g. moving into the same physical space. From a game theory perspective, collective actions are described by joint action profiles rather than individual actions. For example, if an individual pigeon's actions are the directions it can fly, then the flock's action space comprises the possible direction choices of all of the individual birds. 

\section{Constraints on Resource Use: Structure, Coordination and Cooperation}

The structure of a system determines how it can use its computational resources. For example, a single pigeon has a complicated body including sensory organs, neuronal networks, and their wider physiology, which determines how information flows as well as resource constraints. The structure of a flock of pigeons includes all of the complicated structure within each individual as well as the spatial and social relationships between the birds. This structure constrains the system: resources within one bird cannot be directly applied to information held by other birds.

\subsection{Coordination}

The distributed and modular structure of collectives raises distinct coordination challenges:

\begin{itemize}
    \item \textbf{Synchronisation}. Tasks may require specific spatial or temporal ordering \cite{anderson2001teams, malone1994interdisciplinary}. For example, ants collectively transport large objects through synchronised and coherent movements \cite{mccreery2016collective}.  
    \item \textbf{Communication Costs}. Information sharing involves some overhead of costs, i.e. time and energy spent sending and receiving signals \cite{bradbury1998principlesCh9}, which can grow rapidly with group size \cite{brooks1995mythical}. 
    \item \textbf{Individual-Collective Integration}. Dividing tasks into smaller subtasks and then combining results can lead to inefficiencies  depending on the type of task \cite{steiner1972group, malone1994interdisciplinary, anderson2001teams}. For example, precise manipulation of nuts and stones in nut-cracking by chimpanzees \cite{howard2024nonadjacent} cannot easily be shared (and is not a collective activity), while cooperative hunting requires chimpanzees to work together \cite{boesch1989hunting}.   
    \item \textbf{Bandwidth Constraints}. There are limits to how much information individuals can exchange, where signals with higher information density are more prone to error  \cite{bradbury1998principlesCh8, mackay2003information}. 
\end{itemize}

\subsection{Cooperation}

Exacerbating these challenges, collectives are composed of autonomous individuals with potentially conflicting motivations \cite{steiner1972group}. Various forms of conflict can arise, with various mechanisms to resolve conflict. 

Individuals may simply have different preferences (e.g. pigeons who have learnt different routes home). In this case, effective cooperation means resolving these diverging preferences, which may take the form of compromise (aggregation of preferences), leadership (following preferences of a subgroup), or splitting (group fission). In navigation tasks, animal groups across varied taxa elegantly solve this problem through a decision rule based on the angular disagreement in directional preferences: averaging preferences when differences are small and following a leader (or splitting) when differences become too large \cite{biro2006compromise, sridhar2021geometry, couzin2005effective, dyer2009leadership, strandburg2015shared}. 

Collectives also face social dilemmas \cite{olson1965logic} where individuals can free-ride by limiting their contributions while enjoying collective benefits \cite{steiner1972group}. Free-riding can degrade group performance \cite{steiner1972group}; in collective vigilance, selfish groups have lower predator detection rates than cooperative groups \cite{pulliam1982scanning}. An extensive literature explores social dilemmas and mechanisms for cooperation: kin selection favours cooperation between relatives \cite{hamilton1963evolution, axelrod1981evolution}, reciprocity incentivises cooperation in repeated interactions \cite{axelrod1981evolution, trivers1971evolution}, and interdependence considers direct and indirect ways that helping others may ultimately help oneself \cite{roberts2005cooperation, aktipis2018understanding, stewart2024resolving}. How each of these mechanisms apply depends on the problem context, group structure and the nature of interactions \cite{nowak2006five}. 

Interestingly, some collectives elegantly sidestep the problem of cooperation through ``participatory computation'' such that contributions and benefits are combined in the same mechanism. A pigeon cannot completely free-ride on collective navigation; by flying with the flock they are necessarily contributing their input. 

Finally, collective problem solving can emerge incidentally from selfish behaviour. Individuals seeking cover behind each other can amalgamate into a ``selfish herd'' \cite{hamilton1971geometry}, selfish sentinels can provide collective vigilance \cite{clutton1999selfish, bednekoff1997mutualism}, and self-interested trading in markets can support price discovery and the efficient allocation of resources \cite{smith1962experimental, fama1970efficient, hayek2013use}. 

Despite challenges in coordination and cooperation, empirical evidence shows that collectives are able to leverage their greater resources to unlock new ways of solving problems, as we will show in the following section. 

\section{Collective Representations and Algorithms (Algorithmic Level)}

\begin{figure}[ht]
\centering
\includegraphics[width=.9\linewidth]{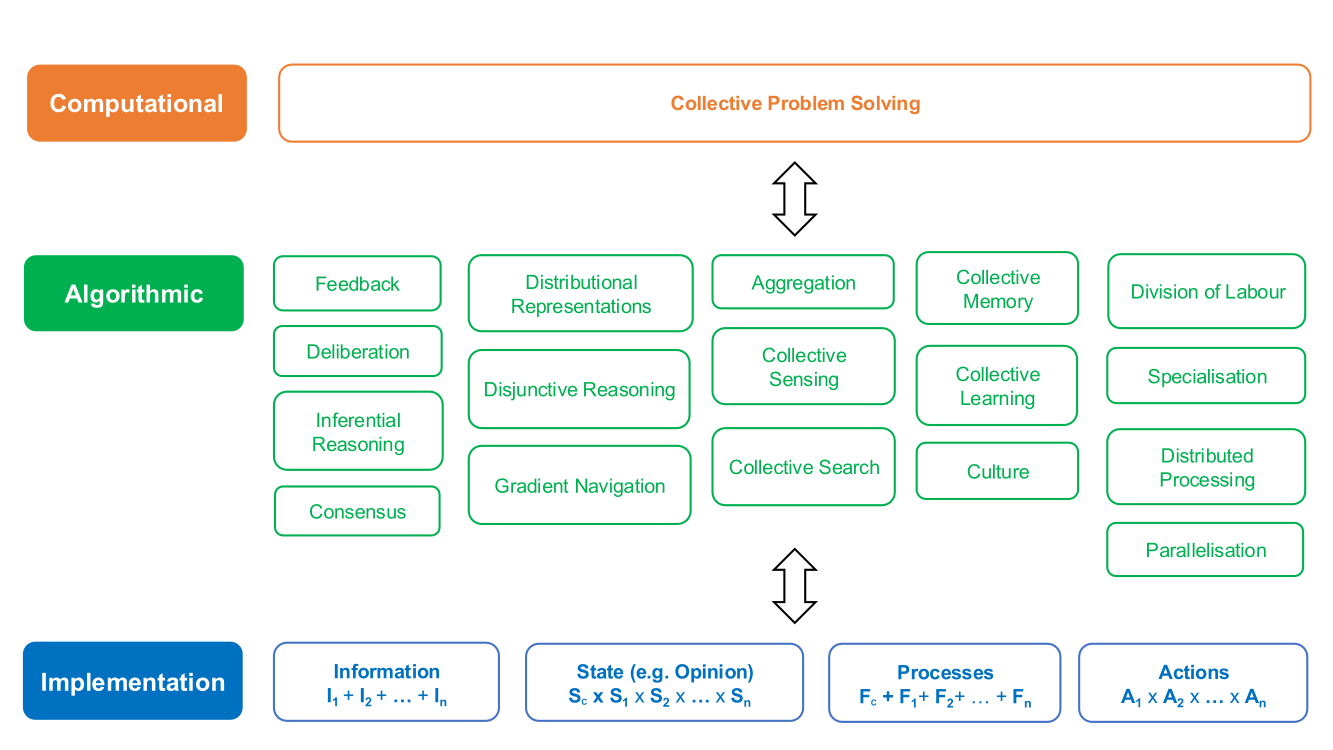}
\caption{Collective resource advantages (bottom) translate to a variety of collective algorithmic capacities (middle) that can be used to solve problems (top).}
\label{fig:collective-algorithms}
\end{figure}

The algorithm level is concerned with how information is represented and how those representations are transformed \cite{marr2010vision}. Our focus will be in analysing how collective physical resources translate to capacities in representing and transforming information (Table \ref{tbl:algorithmic_capacities}), and in doing so connecting those algorithmic capacities to forms of collective intelligence (Figure \ref{fig:collective-algorithms}). 

\subsection{Information Aggregation}

Information aggregation leverages the distributed representation of sensory information, $\sum I_i$, as well as distributed memories or ``belief'' states, $\prod S_i$. Collectives enjoy greater capacities in collective sensing \cite{couzin2009collective, lima1995back, hein2015evolution}, i.e., aggregating sensory information about different things; as well as greater accuracy through the central limit theorem \cite{sasaki2022empirical, surowiecki2005wisdom, francis1907vox, simons2004many, page2008difference, becker2017network, sumpter2006principles}, i.e., aggregating independent information about the same things. 

Collective sensing allows individuals to share information about the environment, such that the collective acts as a distributed sensory network and effectively extends the sensory range of individuals \cite{couzin2009collective, lima1995back, hein2015evolution}. One form of collective sensing is collective vigilance, whereby a group is able to detect predators more efficiently than individuals, employing a many-eyes strategy with all individuals keeping a low-level lookout \cite{pulliam1973advantages, powell1974experimental, magurran1985vigilant, clutton1999selfish}, or coordinating actions, $\prod A_i$, in a sentinel strategy with one (or a few) individuals keeping a dedicated lookout \cite{wickler1985coordination, rasa1986coordinated, gaston1977social, clutton1999selfish}. 

Where the collective contains information about the same things but with uncorrelated errors, the aggregate of the crowd is more accurate than individual opinions \cite{sasaki2022empirical, surowiecki2005wisdom, francis1907vox, simons2004many, page2008difference, becker2017network, sumpter2006principles} (Table \ref{tbl:algorithmic_capacities}), known as the wisdom of the crowd or the many-wrongs principle. Aggregation typically involves combining opinions through consensus mechanisms \cite{simons2004many, becker2017network, seeley2003consensus}, with the most appropriate aggregation algorithm depending on the problem context and informational structure, which may include pockets of correlated information \cite{gray2012modeling, kao2019modular}. Additionally, there can be tradeoffs between accuracy and other concerns such as maintaining information diversity \cite{lorenz2011social, lazer2007network}. 

\begin{table}[ht]
\centering
\small
\renewcommand{\arraystretch}{1.2}
\begin{tabularx}{\textwidth}{@{} l >{\raggedright\arraybackslash}X >{\raggedright\arraybackslash}X @{}}
\toprule
\textbf{Capacity} & \textbf{Individual} & \textbf{Collective} \\
\midrule
Noise, $\sigma^2$          & $\sigma_i^2$                   & $\sigma_i^2/n$ \\
Belief, $H$                & $H_i \in \{a,b,\dots\}$        & $P(H)$ \\
Memory Persistence, $T$    & $T_i \le \text{lifetime}$        & $T_c \le \infty$ \\
Utility Mapping, $u$ & $u(A_i)$ & $u(A_i, A_j)$ \\
\bottomrule
\end{tabularx}
\caption{Examples of algorithmic capacities of individuals and collectives. Where individual opinions have uncorrelated noise, the noise of the collective aggregate opinion scales inversely with the number of individuals, $n$. If individuals each have a set of possible beliefs, then the collective can represent a distribution over this set, $P(H)$. Individual memories can persist only as long as the individual's lifetime, whereas collectives can store information indefinitely by sharing and copying information across the group. Individual actions, $A_i$, can be mapped to expected utilities, while in the collective this mapping is to a $n$-dimensional payoff matrix (in this case 2 dimensional).}
\label{tbl:algorithmic_capacities}
\end{table}

Information aggregation extends to more complicated correlation structures. Environmental features are often spatially autocorrelated, meaning nearby points tend to be similar, forming patches or gradients \cite{legendre1993spatial}. Spatially separated collectives can generate a distributed spatial representation or map of the environment, which effectively exploits this autocorrelation to infer parts of the environment that are not directly observed. This representation can facilitate navigation algorithms in collectives \cite{berdahl2013emergent}, which we will explore in the case of golden shiner fish in the Case Studies section. Somewhat analogously, in computer science, the simplex algorithm navigates gradients by taking a few samples and shifting away from less favourable samples towards more promising directions \cite{nelder1965simplex}. 

\subsection{Distributions and Inference}

The combined states of individuals, $\prod S_i$, form a distributional representation that can capture uncertainty \cite{velez2023humans}. For example, in a flock of homing pigeons, each individual might have a belief about the best direction to travel, in which case the flock contains a distributional representation of beliefs (Table \ref{tbl:algorithmic_capacities}). 

Distributional representations facilitate inferential algorithms that reason about uncertainty \cite{sanborn2016bayesian, griffiths2010probabilistic}. Novel sensory information can be integrated through interactions that amplify or dampen states \cite{sumpter2006principles} across the collective and in doing so alter the distributional representation. This is one path to inferential-style collective algorithms that are able to effectively reason in the Bayesian sense by updating likelihood distributions \cite{krafft2021bayesian, hardy2022overcoming}.

Speculatively, we predict that distributed representations over complex states can unlock other forms of reasoning. Disjunctive reasoning (A ``or'' B) requires a representation of both possibilities, which can be represented collectively even if individuals are only able to represent a single option. Similarly, counterfactual reasoning requires representations of alternative, unrealised outcomes, which can be distributed across individuals in a group. 

\subsection{Feedback and Deliberation}

Inter-individual interactions, $F_C$, allow individuals to influence the state of other individuals. These interactions are a mechanism for feedback, whereby the output of a process is fed back into the process itself. Feedback, which is a focus in control theory \cite{Wiener1961Jan} and self-organisation \cite{sumpter2006principles, Camazine2003Sep}, enables adaptive control of a system and is associated with step-changes in computational capabilities, e.g. combinational logic in circuit design \cite{Mano1990Oct} and Type 2 deliberative reasoning in cognitive science \cite{Kahneman2012May, thompson2009dual}. 

Feedback allows for adaptation of the collective computation to the problem context. Under threat of predation, individual vigilance, $I_i$, increases and inter-individual interactions, $F_C$, become more tightly coupled, boosting responsiveness to weak stimuli \cite{couzin2007collective}. More generally, collectives can adapt their interaction structure, $S_c$, and interaction rules, $F_c$, to varying problem contexts \cite{galesic2023beyond}. 

Collectives can potentially use feedback processes to adaptively control computation in a way analogous to dual process theory (Type 1 and Type 2 reasoning) in psychology \cite{miller2001integrative}. Distributional representations, $\prod S_i$, hold information about the uncertainty within a collective, which in principle can be used to control deliberative processes. When enough individuals agree, there is little benefit in further deliberation and the collective can make a quick decision \cite{sumpter2009quorum}; conversely, if there is discord within a group then this can prompt a slowing down in the decision-making process \cite{strandburg2015shared, couzin2005effective}, analogous to human cognition switching from Type 1 intuitive to Type 2 deliberative reasoning \cite{Pennycook2015Aug}.

\subsection{Collective Memory}

Collective-level state, $S_c$, (i.e., spatial or network structure) can encode information represented in the relationships between individuals as a form of collective memory that can influence future behaviour \cite{couzin2002collective, couzin2009collective}. Collective memory creates the potential for collective learning where inter-individual interactions that lead to favourable outcomes are reinforced, while those that do not are dampened \cite{biro2016bringing}. Empirically, collective learning in the form of adaptive social networks has been shown to boost collective problem solving in human groups \cite{almaatouq2020adaptive}.

The modular structure of collectives includes individuals with their own internal states, $S_i$, which supports a distributed representation of memory across the group, enabling redundant or ``backed up'' copies of information \cite{Chittka2005Nov}. A consequence of redundancy is that information is more resilient to the loss of any single individual's memories \cite{wu2008collaborating}, and information can persist in a collective beyond the lifetime (or membership) of any one individual  \cite{sasaki2017cumulative} (Table \ref{tbl:algorithmic_capacities}). When supported by inter-individual processes, $F_c$, of communication and copying, this memory persistence enables long-term cultural learning and cumulative cultural evolution, whereby knowledge and competencies can build up over generations \cite{boyd1988culture, tennie2009ratcheting, tomasello1993cultural, boyd1996culture}. 

\subsection{Division of Labour and Specialisation}

The collective action space, $\prod A_i$, allows access to strategies not available to individuals, including entirely new ways of solving problems and the exploitation of novel niches. Collective hunting strategies allow a group to work strategically together to capture types of prey that would be impossible for individuals \cite{stander1992cooperative, hansen2023mechanisms, gazda2005division, anderson2001teams}. In the language of game theory, collectives have a joint action profile that maps to expected utilities in a richer way than individual actions (Table \ref{tbl:algorithmic_capacities}).

Taking on complementary roles is a form of division of labour that can boost group performance \cite{smith1776inquiry, becker1992division, anderson2001teams}, such as taking on different tasks within an ant colony  \cite{gordon1996organization}. Individual capabilities in fulfilling roles can be enhanced through specialisation with e.g., in-lifetime learning in humans \cite{campbell1993theory, ericsson1993role} and evolutionary adaptation in e.g., eusocial caste systems \cite{oster1978caste}. This process of specialisation can become irreversible so that the group becomes obligate cooperators; a characteristic of major transitions in individuation such as the emergence of eukaryotes, multicellularity, and eusociality \cite{west2015major, szathmary1995major}.

In general, division of labour is fruitful when a task can be profitably decomposed into subtasks and recombined \cite{malone1994interdisciplinary, steiner1972group}, i.e., parallelised. Complex real-world problems are often decomposable, and parallelisation can be an efficient way of searching for solutions \cite{herbert1962architecture}. Collectives are naturally modular, with tightly coupled processes within individuals, $F_i$, and looser coupled processes between individuals, $F_c$. This structure is well suited to parallelisation, embodying a principle of Marr that efficient representations will mirror the structure of real-world problems \cite{marr2010vision}. 

\section{Case Studies (Computational Level)}

We present case studies that demonstrate the utility of our framework in analysing and describing collective problem-solving behaviour. 

\subsection{Golden Shiners Navigating Light Gradients}

Golden shiner fish are able to navigate as a shoal towards preferred darker regions of their environment much more effectively than individuals \cite{berdahl2013emergent}. Individual fish change their speed depending on the local level of light, going faster in lighter areas. This differential in speeds, combined with inter-individual attraction, causes the entire shoal to turn towards darker areas in a process that resembles phototaxis in plants. This behaviour has been simulated with fish who, individually, have only a local representation of the light level and no representation of the light gradient \cite{berdahl2013emergent}. A similar mechanism has been proposed in collective navigation during animal migrations \cite{couzin2018collective}. 

\begin{figure}[ht]
\centering
\includegraphics[width=.9\linewidth]{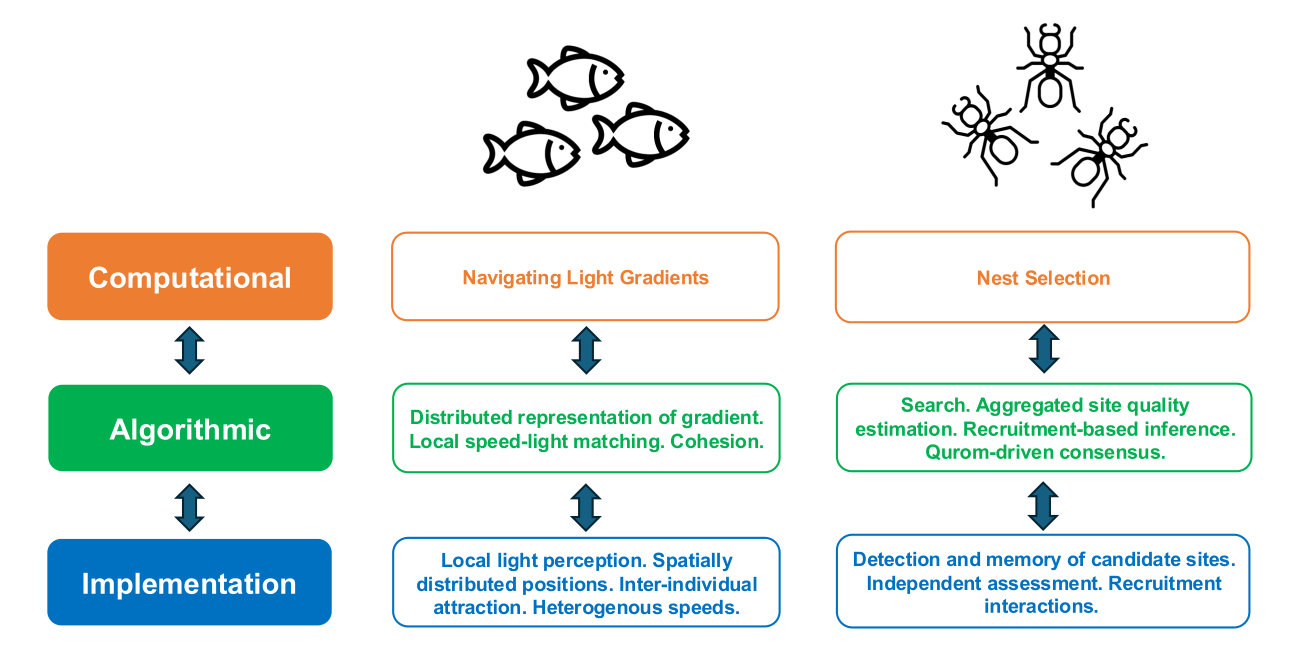}
\caption{Analysis of problem solving with golden shiners navigating light gradients (left) and ants choosing a new next site (right).}
\label{fig:cases}
\end{figure}

Analysed within our framework (Figure \ref{fig:cases}), the real-world problem is navigation towards dark regions. Such navigation requires a representation of the light gradient. While individuals may only have a representation of the local light level, the collective contains a spatially distributed set of samples that generates a representation of the light gradient. This representation presents the opportunity for an algorithm to solve the navigation problem, which is done by leveraging the collective action space by matching heterogeneity in actions (speed of the fish) to the local light level. Inter-individual attraction, which turns the fish shoal, is a feedback process that maintains the shoal density. The interaction of the collective representation of the gradient, coordination in actions to light levels, and feedback processes generate a collective algorithm that navigates towards dark regions in a way that is not accessible to individuals. This mechanism is participatory, so that individual contributions and benefits accrue from the same behaviour, elegantly resolving challenges of coordination and cooperation.

\subsection{Ant Colonies Choosing New Homes}

When ant colonies migrate, they instantiate a well-studied algorithm to find a new nest site  \cite{sumpter2006principles, franks2003strategies, pratt2005agent, franks2002information, sasaki2011emergence, pratt2006tunable}. To begin with, scout ants explore the environment for potential nest sites. When a scout ant finds a good candidate site, they go back to the colony and attempt to recruit other ants to also scout the site, with the vigorousness of their recruitment efforts proportional to their appraisal of the quality of the site. As other ants are recruited, they also assess the nest site, and recruit according to their own independent appraisal. This process happens simultaneously with several candidate nest sites, creating a race dynamic until one site reaches a quorum of a certain number of ants, at which point those ants begin to physically pick up other ants and take them to the new site, leading to rapid consensus throughout the colony. 

Applying our framework (Figure \ref{fig:cases}), the real-world problem is making a fast and accurate decision of a good nest site. The problem can be broken into sequential phases, with an evolving collective representation transformed through different forms of collective intelligence:

\begin{enumerate}
    \item Search for candidate sites. Scout ants explore the environment through collective sensing, generating a distributed spatial representation of candidate sites over a large area. 
    \item Assessment of quality of sites. Recruitment of ants to independently assess nest sites leverages the central limit theorem to generate a representation of the estimated quality of candidate nest sites, instantiated as the intensity of recruitment for each site. This is a feedback process akin to active inference that incorporates new data into the collective representation. The estimation of the relative quality of different sites becomes more accurate as more ants independently assess candidate sites, until a quorum is reached. 
    \item Decision. The colony switches from a slow deliberative process to a rapid consensus mechanism, collapsing the collective distribution of preferences to a single decision. 
\end{enumerate}

Ants coordinate in a distributed way \cite{pratt2006tunable}. A search process begins when the old nest is damaged or disturbed, with individual scouts triggered to explore when locally sensing damage. Quorum sensing is also distributed, based on reaching a threshold of ant encounter rates within a new nest site. Cooperation in ant colonies is supported through high inclusive fitness, such that individual adaptive goals align closely with the success of the colony \cite{bourke1995social}. 

Experiments have compared nest site selection by ant colonies and individual ants. Colonies have been shown to be resistant to a decoy effect that leads to irrational choices in individual ants \cite{sasaki2011emergence, edwards2009rationality}. Colonies outperform individuals when the quality difference is difficult to assess, but notably individuals outperform the colony when the task is easier, with the colony more likely to make a suboptimal choice due to the amplification of noisy assessments early on in the algorithm \cite{sasaki2013ant}, fundamentally arising through the coordination challenge of distributed computation. 

\subsection{Homing Pigeon Collective Navigation}

Homing pigeons move towards their home target in cohesive flocks, performing collective navigation through decision-making algorithms that operate on the directional preferences of individual flock members. These directional preferences can result from pigeons having previously learned different routes home (or arise simply from individual variation), generating a range of route preferences across the flock \cite{biro2006compromise}. However, the directional preferences of individual pigeons are balanced in their output with a drive to fly with other pigeons rather than alone (or in some cases the group splits into subgroups) \cite{biro2006compromise}. The precise outcome reached by a particular flock is influenced by the dynamics of leadership within the flock, which in turn is related to flight speed and personality traits \cite{pettit2015speed, pettit2013interaction, sasaki2018personality}, and is both reflected and implemented through the relative positions of birds within the flock \cite{pettit2013interaction, nagy2010hierarchical}.

The real-world problem that pigeons are solving is efficient navigation home (Figure \ref{fig:pigeon}). Individual pigeons have a range of experiences and memories of the environment, creating a higher-fidelity collective representation of the route home. The challenge for the collective algorithm is to use this distributed knowledge to make decisions. A good algorithm will aggregate directional preferences to leverage the central limit theorem (the many-wrongs hypothesis) and we see evidence for this in greater performance in larger groups \cite{sasaki2022empirical}. Beyond this, an effective aggregation algorithm will be weighted such that birds with greater knowledge contribute more, which we also observe with more experienced pigeons having greater influence \cite{nagy2010hierarchical, pettit2015speed}, with an influence mechanism related to the collective state of the flock \cite{pettit2013interaction, nagy2010hierarchical}. 

\section{Discussion}

Collectives differ from individuals in both their resources (sensing, memory, processing, actions) and constraints (requiring coordination and cooperation). These differences generate distinct collective resource-quality tradeoffs that produce enhanced problem-solving (or collective failures). It is not just that collectives have ``more stuff''. Their modular structure imposes constraints that give rise to emergent information representations and algorithms. From this perspective, diverse behaviours including cultural learning, vigilance, navigation, and cooperative hunting, can be understood as expressions of the same foundational principles. Our case studies demonstrate how the framework applies to specific collective behaviours, including those that combine multiple mechanisms. Overall, our contribution is a step beyond cataloguing examples; we present a general, systematic set of principles of collective intelligence. 

Beyond providing a descriptive language for existing research on collective behaviour, our framework raises concrete questions for future research: 

\begin{itemize}
    \item \textbf{Collective reasoning}. Can collectives exhibit reasoning capacities not available to individuals in the group, such as inferential, disjunctive, and counterfactual reasoning? 
    \item \textbf{Collective biases}. To what extent are collective algorithms `tuned' to the resources, structure and problem being solved: do they collapse or lead to poor outcomes then the context is changed? Do collectives exhibit biases away from rationality? How do these compare to individual biases? Under what conditions do collectives underperform individuals? 
    \item \textbf{Collective intelligence, fast and slow}. Do collectives exhibit fast Type 1 style heuristic decision-making and slower Type 2 style reasoning-like deliberation \cite{Kahneman2012May}? When and how do they switch? Is switching influenced by disagreement within the group? More generally, by what mechanisms do collectives adapt to changing contexts and problems \cite{galesic2023beyond}?
    \item \textbf{Evolutionary pathways}. Historically, did constraints at the resource and structural level drive innovations towards specialisation and cooperation? Was this a factor in major transitions in individuality \cite{west2015major}? How does the expressivity of language facilitate collective algorithms and problem-solving through increasing the bandwidth of interactions \cite{velez2023humans, fusaroli2012coming}? 
    \item \textbf{Representation and perception}. A principle of Marr's is that an effective representation will mirror the structure of the problem. In the context of collectives, does this also work in reverse? Does the perceived modular structure of the world \cite{herbert1962architecture} actually reflect the modular structure of collectives, with human perceptual and linguistic adaptations towards managing reasoning within groups? 
\end{itemize}

Although our focus has been on animal groups, the principles described here apply to other collectives. Human social systems are collectives that perform distributed computations \cite{velez2023humans}, with consequences in our ability to solve global collective action problems such as climate change \cite{adger2010social}, but also for generating harms such as political polarisation \cite{bail2018exposure} and market failures \cite{hardin1968tragedy}. While our framework can help to understand human collective decision making, it can also play a role in designing effective systems, which will be especially important as we integrate artificial intelligence into society. At a smaller scale, parallels with neuronal systems suggest that cognition itself can be understood as a collective process between neurons \cite{couzin2009collective}, supporting the claim that``all intelligence is collective" \cite{falandays2023all}. A benefit of the general framework we present here is that it is compatible with such perspectives, and provides a shared language that applies across collectives at a diverse range of levels of biological organisation \cite{mcmillen2024collective} including genes in genomes, cells in tissues, organs in organisms, species in ecosystems, and humans in society. 

Marr's traditional approach is to begin with the computational problem being solved, and using that framing to ask how a system approximates the ideal computational solution at the algorithmic and implementation levels \cite{marr2010vision}. In this article, we instead began at the implementation level, followed by how resources translate to capacities at the algorithmic level, before bringing it all together at the computational level through case studies. We feel that this was the most natural way to approach our investigation of how computational resources lead to enhanced collective problem-solving. In general, we very much support Marr's top-down approach for investigating specific behaviours. While Marr's levels were developed to analyse individual behaviour, they can be applied to collectives as demonstrated by existing examples in analysing human teams in naval navigation \cite{hutchins1995cognition} and self-organising human social behaviour such as waiting in line \cite{krafft2018levels}. 

The cognitive revolution in psychology took seriously the idea of cognition as computation \cite{newell1994unified, anderson2013architecture}, leading to an explosion of insights and a new field in cognitive science. We invite researchers to take inspiration from that revolution and to take seriously the idea of collective behaviour as computation \cite{velez2023humans}. 

\bibliographystyle{plain}
\bibliography{refs}

\end{document}